\documentclass{CECS}
\title{Path integral measure for first order and metric gravities}
\author{Rodrigo  Aros \\ Universidad Nacional Andr\'{e}s Bello, Sazie 2320, Santiago,
Chile\\E-mail: \email{raros@unab.cl}}
\author{Mauricio Contreras\\ Universidad Nacional Andr\'{e}s Bello, Sazie 2320, Santiago
\\E-mail: \email{mcontreas@unab.cl}}
\author{Jorge Zanelli\\ Centro de Estudios Cient\'{\i}ficos  (CECS), Casilla 1469,
Valdivia, Chile \\E-mail: \email{jz@cecs.cl}}


\abstract{The equivalence between the path integrals for first order gravity and the standard
torsion-free, metric gravity in $3+1$ dimensions is analyzed. Starting with the path integral for
first order gravity, the correct measure for the path integral of the metric theory is obtained.}

\begin{document}

\section{Introduction}

Gravitation as originally formulated by Einstein is a field theory for the
metric. The Einstein-Hilbert action,
\[I_{EH}[g]=\int\sqrt{g^{(4)}}R d^4x,\]
is the only functional of the metric, up to a cosmological constant, whose variation yields second
order field equations for $g_{\mu\nu}$ in four dimensions.

This action can also be written in terms of the local orthonormal basis of the (co)tangent space
(vierbein), $e^{a}$, and the spin connection $\omega ^{ab}$, as
\begin{equation}
I_{EH}[e,\omega ]=\frac{1}{32\pi G}\int_{\mathcal{M}}R^{ab} \wedge e^{c}\wedge e^{d}
\varepsilon_{abcd} \label{EH}
\end{equation}
where $R^{ab}=d\omega ^{ab}+\omega _{\hspace{1ex}c}^{a}\wedge\omega ^{cb}$ is the curvature two
form, related to the Riemann tensor by $R^{ab}= \frac{1}{2}R_{\,\,\,\mu \nu}^{ab}dx^{\mu}\wedge
dx^{\nu}$), which contains only up to first derivatives of the fields.

Extremizing the action yields only first order differential equations for these fields. The
resulting equations are equivalent to Einstein's second order field equations for the metric.
Thus, this first order action describes the same classical system as General Relativity
\cite{Zumino:1986dp,Regge:1986nz,gockeler}. The first order formulation is similar to the Palatini
approach, where the inverse metric does not enter in the action, and the metric and affine
connection are varied independently. In both cases the vanishing of the torsion tensor is not
postulated but is a consequence of the field equations.

The \textit{on shell} equivalence between metric gravity and the first order theory in four
dimensions is easily recognized \cite{Ashtekar:1988sw,Henneaux:1983vi,Henneaux:au}. The purpose of
this note is to establish the off shell equivalence as well by comparing the path integrals for
the first order and the metric formulations of the Einstein-Hilbert theory in four dimensions.

The problem of quantizing the gravitational field has been extensively discussed over the past
seventy years. Different attempts to turn the metric into quantum field have led to uncontrollable
divergences. In the past twenty years some conceptually different ideas have been put forward
(see, e. g., \cite{Isham:1996de}). In some sense, gravity can be viewed as an effective low energy
remnant of the fundamental string \cite{Aharony:1999ti}. Alternatively, spacetime can be construed
as a tapestry made out of fundamental loops roughly $10^{-33}$cm in length
\cite{Gambini:1996ik,Pullin:2002vb}, etc.

Regardless of what the ultimate picture might be, in some approximate sense,
quantum gravity might be represented through a path integral for some
fundamental local field, be it a gauge connection, the vierbein or the metric.
One can then ask whether the different formulations are equivalent to each
other or not. Here we show that the path integral for the first order theory is
formally the same as the one for the metric formalism. Any attempt, however, to
prove or disprove that either formalism is renormalizable --which, by the way,
has not been analyzed in the first order form-- is beyond the scope of this
work.

\section{Metric and first order actions}

In this section we review the expressions of the path integrals for the Einstein-Hilbert theory,
both in the standard metric form (see, e.g., \cite {Teitelboim:ua}) and in the first order
formalism.

\subsection{Second order (metric) gravity}

In the standard metric formulation the torsion is set identically to zero. The Hamiltonian action,
\cite{Dirac:1958,ADM:1962,Henneaux:1992ig} reads
\begin{equation}
I[g ,\pi_g]=\int d^{4}x (\dot{g}_{ij}\pi^{ij}- NH_{\perp }- N^{i}H_{i}),
\label{EHcan}
\end{equation}
where $g_{ij}$ and $\pi^{ij}= G^{ijmn}(g) \dot{g}_{ij}$ are the phase space
coordinates, and
\begin{equation}
G^{ijmn}(g)= g^{1/2} \left[ g^{im} g^{jn} + g^{in} g^{jm}-2g^{ij} g^{mn} \right],
\label{Supermetric}
\end{equation}
is the supermetric (here $g\equiv \det{g_{ij}}$). The Lagrange multipliers $N$
and $N^{i}$ enforce the first class constraints $H_{\perp}\approx 0$ and
$H_{i}\approx 0$, respectively. There are 6 q's, 6 p's, which together with the
4 first class constraints yield 2 propagating degrees of freedom.  The path
integral reads
\begin{equation}
Z_{g} =\int [Dg_{ij}][D\pi^{ij}] [D N][D N^{i}] \exp \left[\frac{i}{\hbar}
I[g,\pi_g]\right]\times [Ghosts], \label{PathIntegral(g)}
\end{equation}
where ``$[Ghosts]$" represents the measure for the ghost and antighosts needed to fix the
diffeomorphism invariance of the theory. This measure has been extensively discussed in the
literature and different proposals have been advanced
\cite{Teitelboim:ua,Fradkin:df,Fujikawa:1984qk}. We shall not consider this contribution to the
path integral measure here. Since both the metric and the first order theories share the same
invariance under diffeomorphisms, we shall assume that, for a given prescription for $[Ghosts]$ in
one formulation there is a corresponding equivalent in the other.

After integrating out the momenta $\pi^{ij}$, Eq.(\ref{PathIntegral(g)}) can be
rewritten (modulo ghosts terms) as
\begin{equation}
Z_{g} =\int \frac{[Dg_{ij}][DN][DN^{i}]}{\sqrt{\det( N G_{ijmn})}} \exp
\left[\frac{i}{\hbar} \int_{{M}} \sqrt{g^{(4)}}R\mathit{\ }d^{4}x\right].
\label{PathIntegral(g)2}
\end{equation}

\subsection{First order gravity}

Two descriptions of first order gravity can be considered, the $e$-frame and
the $\omega$-frame \cite{Banados:1997hs,Contreras:1999je}. Both frames are
related by a canonical transformation and therefore have the same classical
action, modulo boundary terms. In the $e$-frame the fields are the vierbein
$e^a_k$ and its conjugate momentum $P^k_a$, and only first class constraints
are present. In the $\omega$-frame the action is a functional of the spin
connection and its conjugate momentum, and both, first and second class
constraints appear. In \cite {Contreras:1999je}, both frames were shown to be
quantum mechanically equivalent.

The field equations obtained extremizing Eq.(\ref{EH}) with respect to $e^{a}$ are
\begin{equation}
\varepsilon _{abcd}R^{bc}\wedge e^{d}=0,  \label{Einstein}
\end{equation}
which are equivalent to the usual Einstein equations. Varying with respect to
$\omega _{ab}$, yields \footnote{In the variation leading to (\ref{Torsion}) an
integration by parts was performed. This usually brings in a boundary term
which here is assumed to vanish by virtue of some appropriate boundary
conditions. This is the case in asymptotically flat spacetimes, or more
generally, if $\omega$ is held fixed at infinity.}
\begin{equation}
\varepsilon _{abcd}T^{c}\wedge e^{d}=0.  \label{Torsion eq}
\end{equation}
Here
\begin{equation}
T^{a}=de^{a}+\omega_{\hspace{1ex} b}^{a} e^{b}, \label{Torsion}
\end{equation}
is the torsion two-form, related to the torsion tensor
($T^{a}=\frac{1}{2}T_{\mu \nu }^{a}dx^{\mu}dx^{\nu }$). Note that, as in the
Palatini formalism, Eq. (\ref{Torsion eq}) implies $T^{a}=0$. In the metric
formalism, instead, $T^{a}$ is assumed to be identically zero, and
$\omega^{ab}$ is not assumed to be an independent field. This means that
although the two formalism give the same classical equations, they need not
define equivalent quantum theories.

\section{The $e$ frame}
In coordinates $(t,x^{i})$, the canonical action in 3+1 dimensions in $e$-frame
reads \cite{Contreras:1999je}
\begin{equation}
I[e ,P_e]=\int d^{4}x (\dot{e}_{k}^{a}P_{a}^{k}-\omega _{t}^{ab}J_{ab}-
NH_{\perp } - N^{i}H_{i}).
\end{equation}
Here $e^a_j$ is the canonical coordinate and its conjugate momentum is
\begin{equation}
P^j_d := \Omega_{d\;ab}^{\;j\;\;k}\; \omega_{k}^{ab},
\label{Momentum-e}
\end{equation}
where $\Omega$ is the symplectic form
\begin{equation}
\Omega_{d\;ab}^{\;j\;\;k}\ \equiv 2\varepsilon_{abcd} \epsilon^{ijk}
\omega_{k}^{ab} e_{i}^{c}.
\end{equation}
The Lagrange multipliers $\omega_{t}^{ab}$, $N$ and $N^{i}$ correspond to the
the first class generators of Lorentz transformations and diffeomorphisms,
respectively. In this frame, the phase space has 12 $q$s and 12 $p$s, there are
10 first class (and no second class) constraints. This gives again 2
propagating degrees of freedom, as in the metric formalism. The resulting path
integral in this frame is therefore given by
\begin{equation}
Z_{e} =\int [De^{a}_{k}][DP_{a}^{k}] [D\omega_{t}^{ab}] [D N][D N^{i}]
\det(M_{\alpha \beta}) \exp \left[
\frac{i}{\hbar } I[e ,P_e], \right] \label{PathIntegral(e-frame)}
\end{equation}
where $M_{\alpha \beta}$ is the matrix of Poisson brackets
\begin{equation}
M_{\alpha \beta}=\{F_{\alpha},\varphi_{\beta}\}^{*},
\end{equation}
where $F_{\alpha}=(\sigma_{\perp},\sigma_{i},\sigma_{ab})$ are the gauge fixing
conditions for the first class constraints $(H_{\perp},H_{i},J_{ab})$
respectively.

In the $\omega$-frame there are 18 $q$s and 18 $p$s, there are also 10 first
class and 12 second class constraints, which also yields 2 propagating degrees
of freedom as well. It was shown in \cite{Contreras:1999je} that the path
integrals in the two frames are equal,
\begin{equation}
Z_{\omega}= Z_{e}=Z_{First\hspace{1ex} Order}.
\end{equation}

\section{Field redefinitions}
It is clear from (\ref {Momentum-e}) that the momentum $P^i_a$ is essentially
proportional to the spin connection. The connection contains a part which is
determined by the vierbein, and a torsion-dependent part. This means that, in
the first order formulation, the torsion tensor is a function of the momentum
canonically conjugate to the vierbein. Thus, the first step to establish the
relation between the path integral of the first order theory
(\ref{PathIntegral(e-frame)}) and that of the metric form (\ref
{PathIntegral(g)2}), can be to separate the metric from the nonmetric
(torsional) components of the spin connection. The torsion tensor
(\ref{Torsion}) can be solved for $\omega^{ab}$, expressing the spin connection
as
\[
\omega _{\mu }^{ab}=\bar{\omega}_{\mu }^{ab}(e)+K_{\mu }^{ab}(e,T),
\]
where $K_{\mu }^{ab}$ is the contorsion tensor and $\bar{\omega}$ is a
torsion-free connection, that is,
\[
de^{a}+ \bar{\omega}_{\hspace{1ex} b}^{a} \wedge e^{b}=0.
\]
Consequently, the momentum $P_c^j$ can be decomposed into a term depending on
the vierbein and a projection of contorsion as $P_c^j = \Omega
_{c\;ab}^{\;j\;\;i}\; (\bar{\omega}_{i}^{ab}(e)+K_{i}^{ab}(e,T) )$, which can
also be written as
\begin{equation}
P_c^j = \Omega _{c\;ab}^{\;j\;\;i}\; \bar{\omega}_{i}^{ab}(e)+K_{c}^{j}(e,T),
\label{pe}
\end{equation}
where all the torsional dependence is contained in the new canonical momenta
$K_a^i$. Since $\bar{\omega}$ has vanishing Poisson bracket with $e$, the
canonical measure in the $e$-$P_e$ phase space can be directly expressed in the
$e$-$K$ space as
\begin{equation}
[De_i^a][DP_a^i]=[De_i^a][DK_a^i]. \label{Measure-E-K}
\end{equation}

On the other hand, decomposing the frame basis $e^a_{\mu}$ along the spatial
directions $e^a_j$ and the timelike normal $\eta_a$, the Lagrange multipliers
$N$ and $N^{i}$ can be written as
\begin{eqnarray}
N(e)&=& \eta _{a}e_{0}^{a}\\
N^{i}(e)&=& E_{a}^{i}e_{0}^{a},
\end{eqnarray}
where $\eta_a \eta^a=-1$, $\eta_a e^a_i =0$, and $E^i_a e^a_j =\delta^i_j$. Now the measure $[D
N][D N^{i}]$ can be shown to be $-\sqrt{g^{(3)}}[D{e^{a}_{0}}]$, where $g^{(3)}= \det(g_{ij})$ and
$g_{ij}=e^{a}_{i} e^{b}_{j}\eta_{ab}$. Thus, the integration measure in
(\ref{PathIntegral(e-frame)}) reads
\begin{equation}
[De^{a}_{k}][DP_{a}^{k}][D\omega_{t}^{ab}][D N][D N^{i}]
=-\sqrt{g^{(3)}}[De_i^a][De_{0}^a][DK_a^i][DK^{ab}_{0}], \label{Measure-N-N}
\end{equation}
where $\omega_{t}^{ab}\equiv K^{ab}_{0}$. In these new variables, the angular momentum is
\[
J_{ab}= K_{a}^{j}e_{b j}-K_{b}^{j}e_{a j}.
\]
The 12 components of $K^{i}_{a}$ can be projected also along spatial and normal directions,
\begin{equation}
K_{a}^{i}=K^{i}\eta _{a}+ \left(\kappa^{(ij)}+ \kappa^{[ij]}\right)e_{aj},
\label{KiKij}
\end{equation}
where $\kappa^{(ij)}$ is symmetric and $\kappa^{[ij]}$ antisymmetric. The
angular momentum constraint can be written in terms of $K^i$ and
$\kappa^{[ij]}$ only, as
\begin{equation}
J_{ab}=K^{i}(\eta _{a}e_{bi}-\eta _{b}e_{ai})+ \kappa^{[ij]}(e_{aj}
e_{bi}-e_{bj}e_{ai}).  \label{Jab}
\end{equation}
It is straightforward to invert this relation, writing $K^{i}$ and
$\kappa^{[ij]}$ in terms of $J_{ab}$,
\begin{eqnarray}
K^{i}&=&-\eta^{a}E^{bj}J_{ab}=-g^{ik}\eta^{a}e_{k}^{b}J_{ab}  \nonumber \\
\kappa^{[ij]}&=&E^{ai}E^{bj}J_{ab}=g^{ip}g^{jq}e_{p}^{a}e_{q}^{b}J_{ab} .
\label{Lab-To_transformations}
\end{eqnarray}

The EH action, written in terms of these fields, becomes
\begin{equation}
\bar{I}=\int_{{M}} \left(\sqrt{g^{(4)}}R - K_{0}^{ab}J_{ab} - N\kappa^{(ij)}G_{ijmn}\kappa^{(mn)}
+ f(J_{ab}) \,\right)d^{4}x, \label{Action-K-e}
\end{equation}
where $G_{ijmn}$ is the inverse of the supermetric defined in (\ref{Supermetric}) and $f(J_{ab})$
is a functional which vanishes for $J_{ab}=0$. In terms of these new fields, the measure in the
$e$-$K$ space becomes
\begin{equation}  \label{KuponK}
[De_i^a][DK_{a}^{i}]= \det(\mathcal{D})[De_i^a][DJ_{cd}][D\kappa^{(mn)}],
\end{equation}
where $\mathcal{D}$ is the Jacobian matrix
\begin{eqnarray}
\mathcal{D}^{\mathbf{M}}_{\mathbf{N}} &=& \left[\partial
K_{a}^{i}/\partial\left(J_{cd}, \kappa^{(mn)}\right) \right] \nonumber \\
&=& \left[ g^{ip}g^{jq} e^{[c}_{p} e^{d]}_{q} e_{aj}- \eta^{[c}e^{d]}_{k}g^{ik}\eta_{a},
\frac{1}{2} (\delta^{i}_{m} e_{a n}+\delta^{i}_{n} e_{a m})\right]
\end{eqnarray}
where the indexes stand for the 12 combinations $\mathbf{M} = [_{a} ^{i}]$ and $\mathbf{N}=
[^{[cd]},_{(mn)}]$.  It is straightforward to show that $\det(\mathcal{D}) = (Const)$, which can
be confirmed by observing that under $e^{a}_{j}\rightarrow \lambda e^{a}_{j}$, determinant of
$\mathcal{D}$ remains unchanged.

In order to make contact with the path integral in the second order
formulation, (\ref{PathIntegral(g)2}), three more steps are in order: first,
integrating over the Lagrange multiplier $K^{ab}_{0}$ in (\ref{Action-K-e})
produces a $\delta (J_{ab})$ which makes $f(J_{ab})$ drop out from the action
and eliminates the integration over $J_{cd}$. Second, integrating over
$\kappa^{(ij)}$ yields a Gaussian form and brings down a factor $[\det
(NG_{ijmn})]^{(-1/2)}$.

Finally, since Lorentz symmetry is not present in the metric theory, it should
be freezed out of the 16 components of $e^a_{\mu}$, replacing them by their
Lorentz invariant components (metric) and six Lorentz rotation coefficients.
This can be done expressing the vierbein in the form
\begin{equation}
e^{a}_{\mu} = U(x)^{a}_{\hspace{1ex} b}\,\hat{e}^{b}_{\mu},  \label{Split-e}
\end{equation}
where $\hat{e}^{b}_{\mu}$ is a fixed vierbein and $U(x)^{a}_{\hspace{1ex} b}$
corresponds to a local Lorentz transformation. Here we shall assume that
$e^a_{\mu}$ is globally defined at least in each spatial section
$\Sigma_{t=t_0}$. This means that there is a gauge (choice of
$U(x)^{a}_{\hspace{1ex} b}$) such that the vierbein is equal to
$\hat{e}^a_{\mu}$ throughout $\Sigma_{t=t_0}$.

In terms of the group parameters $\lambda^{ab}=-\lambda^{ba}$ the local Lorentz
rotations read
\[
U^a_{\hspace{1ex}b} = \delta^a_{\hspace{1ex}b} + (\lambda
\textbf{L})^a_{\hspace{1ex}b} + \frac{1}{2}(\lambda
\textbf{L})^{a}_{\hspace{1ex}c}(\lambda \textbf{L})^c_{\hspace{1ex}b}  +
\ldots,
\]
where $\textbf{L}$ are the generators of $SO(3,1)$ in the vector
representation, expressed as $\left(\textbf{L}_{cd}\right)^{a}_{\hspace{1ex} b}
= \eta_{cb} \delta^{a}_{d}-\eta_{db} \delta^{a}_{c}$. Thus, the 16 components
$e^{a}_{\mu}$ are described by 10 fields corresponding to the rotational
invariant part of the representative vierbein $\hat{e}^{b}_{\mu}$, which can be
identified with the metric $ g_{\mu\nu}=
\eta_{ab}\hat{e}^{a}_{\mu}\hat{e}^{b}_{\nu}$, and the 6 variables
$\lambda^{ab}$.

Varying the expression (\ref{Split-e}) with respect to $\lambda^{cd}$ and
$g_{\mu \nu}$ yields
\begin{equation}
\delta e^a_{\mu}= \frac{\delta U^a_{\hspace{1ex}b}}{\delta \lambda^{cd}}E^{b \nu}g_{\mu\nu}
\delta\lambda^{cd} + U^{a}_{\hspace{1ex}b} E^{b\nu} \delta g_{\mu\nu}.
\end{equation}
The measure of integration over the vierbein can be written as
\begin{equation}  \label{Measure}
[De^{a}_{\mu}] = [D(\lambda^{cd})][ D(g_{\alpha \beta}) ]\det{\mathcal{B}},
\end{equation}
where $\mathcal{B}$ stands for $16\times16$ matrix
\begin{equation}  \label{B}
\mathcal{B}^{\mathbf{M}}_{\mathbf{N}} = \tilde{E}_{b}^{\hspace{1ex} \nu}\left[
g_{\mu\nu} \frac{\delta U^{ab}}{\delta \lambda^{cd}}\hspace{1ex}, U^{ab}
\frac{1}{2}\left(\delta^{\alpha}_{\nu}\delta^{\beta}_{\mu}+
\delta^{\beta}_{\nu}\delta^{\alpha}_{\mu}\right)\right]
\end{equation}
where the indexes are the 16 combinations $\mathbf{M} = [^{a} _{\mu}]$ and
$\mathbf{N}= [_{[cd]},^{(\alpha\beta)}]$.

The assumption that (\ref{Split-e}) be globally defined implies that diffeomorphisms and Lorentz
rotations can be performed independently. Consequently, one can fix the Lorentz frame by choosing
$U^a_{\hspace{1ex}b}= \delta^a_{\hspace{1ex}b}$ globally, so it is always possible to select the
gauge condition by fixing $\lambda=0$ everywhere, say. This yields
\[
\left.\frac{\delta U^{ab}}{\delta \lambda^{cd}}\right|_{\lambda=0} =
\delta^{ab}_{cd}.
\]
which implies
\[
\left.\mathcal{B}^{\mathbf{M}}_{\mathbf{N}}\right|_{\lambda=0} = \frac{1}{2}
\tilde{E}_{b}^{\hspace{1ex}\nu}\left[g_{\mu\nu} \delta^{ab}_{cd}\hspace{1ex},
\eta^{ab} \left(\delta^{\alpha}_{\nu}\delta^{\beta}_{\mu}+
\delta^{\beta}_{\nu}\delta^{\alpha}_{\mu}\right)\right].
\]

It is straightforward to check from (\ref{B}) that $\det(\mathcal{B})= (const)\times
\left(g^{(4)}\right)^{-1/2}$. Thus, the path integral (\ref {PathIntegral(e-frame)}) of the first
order theory can finally be written, up to a multiplicative constant, as
\begin{equation}
Z_{First\hspace{1ex}Order}=\int \frac{D[g_{ij}][D N][D N^{i}]}{ \sqrt{\det
(NG_{ijmn})}}\times \exp \left[ \frac{i}{\hbar} \int_{{M}}
\sqrt{g^{(4)}}R\mathit{\ }d^{4}x\right], \label{PathIntegral-g}
\end{equation}
which coincides with the second order expression, as expected (here we have
used the fact that $\det(\mathcal{D})= (constant)$).

\section{Summary and prospects}

We have shown that the path integral for the Einstein-Hilbert action in four
dimensions and with vanishing cosmological constant is formally identical for
the first order (vierbein) formulation as for the second order (metric) theory.
However, as the steps of the proof depend critically on several features
peculiar, to the four dimensional EH action, it is likely to fail in more
general settings.

\textbf{LL theories.} For spacetime dimensions $D>4$ there exist a family of sensible theories,
including higher powers of curvature but no explicit torsion in the action, with second order
equations for the metric, that generalize General Relativity. These are the so-called
Lanczos-Lovelock theories \cite{Lanczos:1975su,Lovelock:1971}. For them there are also two
versions: a first order and a metric one. But, unlike the EH case, these two formulations are not
classically equivalent for every field configuration. In particular, there are configurations in
which the first order formulation might allow nonvanishing classical torsion
\cite{Troncoso:1999pk}, whereas the second order version always assumes zero torsion. However, the
configurations where the two theories are classically inequivalent form a set of measure zero in
the space of solutions and may be ignored in generic backgrounds.

\textbf{Torsion as momentum.} In the second order formulation, the torsion is assumed to be
identically zero and therefore never varied in the action or integrated over in the path integral.
In contrast, we observe that in the first order approach, torsion is not only allowed to vary, it
is necessary since it represents the canonical momentum conjugate to the vierbein (c.f. Eq.
(\ref{Measure-E-K})).

\textbf{Degrees of freedom.} The off shell equivalence is probably still true
for the LL gravity theories. This is because the torsion-free LL theories have
the same degrees of freedom as the standard Einstein-Hilbert system (see e.g.,
\cite{Teitelboim-Zanelli:87}). On the other hand, the proof is unlikely to go
through for more general actions explicitly involving torsion in the
Lagrangian. Theories of this type were discussed in \cite{Mardones:1991qc}, and
were shown to possess more degrees of freedom in general
\cite{Banados:1995mq,Banados:1996yj,Troncoso:1999pk}.

\textbf{Adding a cosmological constant.}As mentioned above, the boundary terms
that arose in calculations vanished provided a flat asymptotic conditions are
assumed. This implies that the proof remains valid for the theory defined on an
open domain in the presence of a cosmological constant. However, the case
$\Lambda \neq 0$, where the boundary terms at infinity could diverge, should be
analyzed more carefully to see whether a similar picture as that for the
asymptotically flat case can be drawn.

\textbf{The character of the equivalence.} The proof of quantum equivalence presented here is in
any case formal. Principally, because the equality is between two expressions which no one knows
how to unambiguously evaluate, interpret or use to predict any experiment.

In view of this plethora of possibilities it would be interesting to extend
this work, establishing the path ordered integral to some these alternative
theories of gravity.

\acknowledgments

This work is partially funded by grants 1020629, 1010450 and 7020629 from
FONDECYT and grant DI 08-02 (UNAB). The generous support of Empresas CMPC to
the Centro de Estudios Cient\'{\i }ficos (CECS) is also acknowledged. CECS is a
Millennium Science Institute and is funded in part by grants from Fundaci\'on
Andes and the Tinker Foundation.

\providecommand{\href}[2]{#2}\begingroup\raggedright

\end{document}